\documentclass{article}

\usepackage[final]{neurips_2021}

\usepackage[utf8]{inputenc} 
\usepackage[T1]{fontenc}    
\usepackage{hyperref}       
\usepackage{nicefrac}       

\usepackage{graphicx}
\title{Banknote Recognition for Visually Impaired People (Case of Ethiopian note)}

\author{%
  Nuredin Ali \\
  Department of Information Systems \\
  Mekelle University \\
  \texttt{nuredi2000@gmail.com} \\
}

\begin{document}

\maketitle

\begin{abstract}
Currency is used almost everywhere to facilitate business. In most developing countries, especially the ones in Africa, tangible notes are predominantly used in everyday financial transactions. One of these countries, Ethiopia, is believed to have one of the world’s highest rates of blindness (1.6\%) and low vision (3.7\%). There are around 4 million visually impaired people; With 1.7 million people being in complete vision loss. Those people face a number of challenges when they are in a bus station, in shopping centers, or anywhere which requires the physical exchange of money. In this paper, we try to provide a solution to this issue using AI/ML applications. We developed an Android and IOS compatible mobile application with a model that achieved 98.9\% classification accuracy on our dataset. The application has a voice integrated feature that tells the type of the scanned currency in Amharic, the working language of Ethiopia. The application is developed to be easily accessible by its users. It is build to reduce the burden of visually impaired people in Ethiopia. 
\end{abstract}

\section{Introduction}
Currency is used almost everywhere in the world to facilitate business. Although electronic commerce is emerging as an alternative means of making business, paper currency is still an indispensable part of everyone’s daily routine. When it comes to the developing countries, especially African countries like Ethiopia, most of the transactions are made through a tangible note. There are various currencies all over the world, and each of them looks different and unique in size, color and pattern. 
\cite{sarfraz2015intelligent} A system for the recognition of paper currency is one kind of intelligent system which is a very important need of the current automation systems. The automation includes aiding visually impaired people. \cite{alene2019ethiopian} The use of automatic methods of currency recognition has been increasing due to its importance in many sectors. Developing simple applications of AI and computer vision which are best fitted for  the visually impaired members of developing communities is an important application. 

\cite{cherinet2018prevalence} Ethiopia is believed to have one of the world’s highest rates of blindness (1.6\%) and low vision (3.7\%). According to, \url{https://www.together-et.com/} there are about 4 million visually impaired people in Ethiopia of which 1.7 million are completely without vision. Those people face a number of challenges in their day to day activities. For instance, when they are in a bus-station, in shopping centers, or anywhere that requires an exchange of currency they face difficulties because they don't know the currency in their hand. On account of that fact, we built an android and IOS compatible  mobile application with a voice feature integrated that notifies the user when the application is opened. When an image of the note is taken, the application generates a voice in Amharic, the working language of Ethiopia, telling the user the amount of birr (Ethiopian Banknote). The main aim of this application is to enable visually impaired people to know the currency they hold through an easily accessible way and make them part of the market from which they had been marginalized.

\section{Background and related works}
\label{gen_inst}
 The recognition of banknotes has  been  addressed  by ~\cite{alene2019ethiopian} and ~\cite{tessfaw2018ethiopian} in several  ways  depending  on  their  types. To the best of our knowledge, there is no extensive research done on the recognition of Ethiopian birr notes. ~\cite{zeggeye2016automatic} The aim of currency recognition system is to teach machines to recognize different currencies and thereby help people work with convenience and efficiency. ~\cite{alene2019ethiopian} Experimental result shows that convolutional neural networks (CNNs) provide better results in classifying Ethiopian banknote denomination. ~\cite{tessfaw2018ethiopian} shows that, after features are extracted, it is important to recognize the currency using an effective classifier called Support vector machine. These previous works tried to recognize the Ethiopian banknote through different mechanisms. However, As of 2020, Ethiopia has released a set of new banknotes.
 
 In this work, the MobileNet architecture is used for recognising these new banknote. The main aim of this work is to classify the banknote, not for counterfeit detection. ~\cite{dinku2009counterfeit} It is very difficult to discriminate the fake notes from the genuine ones by simply looking at the paper notes.

\section{Data and Methodology}
\label{headings}

\subsection{Dataset}
\begin{figure}[ht]
  \centering
  \includegraphics[width=\textwidth]{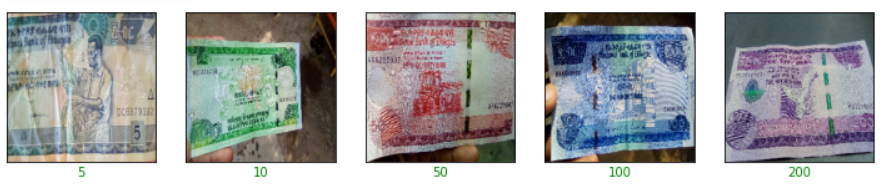}
  \caption{Sample of the data set used}\label{fig 1:}
\end{figure}

This work is conducted on the 2020, newly released, birr notes. There are five different notes 5, 10, 50, 100 and 200, from 10 – 200 being released in a new design and format. \autoref{fig 1:} shows samples of the data that is used in this experiment. A total of 2060 images are used to train the model. The dataset is partitioned into training, validation and test sets. 70\% for training, 15\% for validation and 15\% for testing. The dataset has six different classes: the five type of notes and the "other" class. The "other" class contains images of different objects that are similar to the Ethiopian currency.  Each class contributes 350 images.

\subsection{Methodology}
~\cite{alene2019ethiopian} To classify the Ethiopian banknote, CNN as a feature extraction is the ideal solution. A Convolutional Neural Network algorithm is a multi-layer perceptron that is specially designed for the identification of two dimensional data such as images. Applying a CNN algorithm offers an advantage over traditional machine learning algorithms because it avoids an explicit feature extraction. In other words, it implicitly learns features from the training set by extracting high-level features (sophisticated features). In case of low resource dataset, \cite{yaeger1996effective} Data augmentation, a regularization scheme that artificially inflates the dataset by using label preserving transformations to add more invariant examples, is crucial to improve models' performance. The following augmentation settings are applied to the training set to increase the data: rotation range of 0.2, zoom range of 0.1, width shift range of 0.2, height shift range of 0.2. 

We used MobileNet, an ImageNet pre-trained model for training. As a baseline, all the layers of the pre-trained model are frozen. Since the state-of-the-art MobileNet model was initially trained on data different than banknotes, the results were not satisfactory. We also used the MobileNet architecture by unfreezing all the layers. This highly improved the prediction accuracy of the model as the model learned most of the features of the banknote. 

For all experiments, the following experimental settings were added: Max and average pooling, dense and dropout layers. In addition, a very low learning rate of 0.00001, batch size of 32,  loss function of categorical cross-entropy, softmax activation and Adam optimizer were used to train the banknote recognition model.

An android and IOS mobile application was developed to make the application easily accessible. We used Tensorflow-lite, to integrate the model with mobile devices. The application has an Amharic language integrated that tells what currency is recognized.

\section{Results}

The evaluation metric used in this work is accuracy. For each image, we check if the correct label is found. The model developed by freezing all the layers achieved 30\% testing accuracy. While the model developed by unfreezing the MobileNet architecture achieved a 98.9\% testing accuracy. Figure 2 shows the confusion matrix of the 98.9\% accuracy model on the test set which is also used in the application. The application works fine when tested live, even at recognizing unknown objects. 

\begin{figure}[ht]
  \centering
  \includegraphics[width=0.5\textwidth]{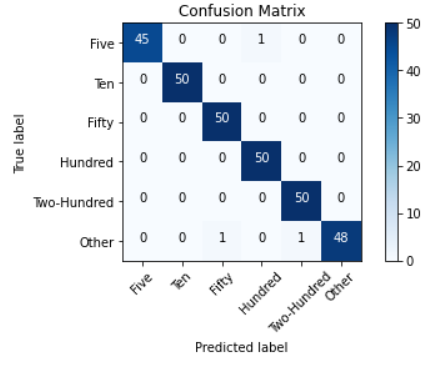}
  \caption{shows the confusion matrix of the model}\label{fig 2:}
\end{figure}

\section{Conclusion and Future work}
In this work, we developed an android and IOS compatible mobile application for visually impaired people that recognizes and tells the currency in hand. We used an ImageNet pre-trained CNN model and achieved a 98.9\%  accuracy.  
In future works, We will cooperate with the Ethiopian National Association of the Blind (ENAB) to test the usability of the application and work on the feedback of the real users. Our goal is to make the life of the visually impaired better using AI/ML applications.


\bibliographystyle{unsrtnat}  

\bibliography{references}

\end{document}